\documentclass[twocolumn]{jpsj2} 
\usepackage{multirow}
\usepackage{array}
\newcommand{\ce}{CeCu$_2$Si$_2$}
\newcommand{\cege}{CeCu$_2$Ge$_2$}
\newcommand{\rz}{$\rho_0$}
\newcommand{\pv}{$P_{\rm v}$}
\newcommand{\pc}{$P_{\rm c}$}
\newcommand{\tn}{$T_{\rm N}$}
\newcommand{\sg}{superconducting}
\newcommand{\sy}{superconductivity}
\newcommand{\tc}{$T_{\rm c}$}
\newcommand{\imgpath}[1]{#1}
\makeatletter\def\@citess#1{\textsuperscript{#1)}}
\title{Valence Instability and Superconductivity in Heavy Fermion Systems}

\author{Alexander T. \textsc{Holmes}$^{1}$\thanks{E-mail address: alex@djebel.mp.es.osaka-u.ac.jp}, Didier \textsc{Jaccard}$^{2}$ and Kazumasa \textsc{Miyake}$^{3}$}

\inst{$^{1}$KYOKUGEN, Center for Quantum Science and Technology under Extreme Conditions, Osaka University, Toyonaka, Osaka 560-8531 \\
$^{2}$D\'{e}partement de Physique de la Mati\`{e}re Condens\'{e}e,
Section de Physique, University of Geneva, 24 quai Ernest-Ansermet\\
CH-1211 Gen\`{e}ve 4, Switzerland\\
$^{3}$Department of Materials
Engineering Science, Graduate School of Engineering Science, Osaka
University, Toyonaka, Osaka 560-8531 }

\abst{Many cerium-based heavy fermion (HF) compounds have
pressure-temperature phase diagrams in which a superconducting
region extends far from a magnetic quantum critical point. In at
least two compounds, \ce\ and \cege, an enhancement of the
superconducting transition temperature was found to coincide with
an abrupt valence change, with strong circumstantial evidence for
pairing mediated by critical valence, or charge transfer,
fluctuations. This pairing mechanism, and the valence instability,
is a consequence of a $f-c$ Coulomb repulsion term $U_{fc}$ in the
hamiltonian. While some non-superconducting Ce compounds show a
clear first order valence instability, analogous to the Ce
$\alpha-\gamma$ transition, we argue that a weakly first order
valence transition may be a general feature of Ce-based HF
systems, and both magnetic and critical valence fluctuations may
be responsible for the superconductivity in these systems.}

\kword{heavy fermion superconductors, mixed valence compounds,
CeCu$_2$Si$_2$, CeCu$_2$Ge$_2$, superconductivity, fluctuations in
superconductors, critical valence fluctuations, new paradigm}

\begin{document}
\maketitle
\section{Introduction}
A resurgence of interest in heavy fermion (HF) superconductivity
in the last few years has been strongly driven by the discovery of
a large number of new Ce and U based intermetallic
superconductors. Most analyses of their behaviour have focused on
the relationship between superconductivity and magnetism,
specifically on the idea that superconductivity in these systems
is mediated by low energy magnetic fluctuations around a so-called
magnetic quantum critical point (QCP), where a magnetic ordering
temperature is driven to zero by an external parameter such as
pressure or chemical substitution.

The aim of this article is to draw attention to another phenomenon
in Ce compounds, namely a weakly first order valence instability
also capable of generating superconductivity. In this case the
pairing is mediated by the exchange of critical valence, or
charge-transfer, fluctuations (CVF)\cite{Onishi00}. It is the view
of the authors that this scenario is of widespread importance, and
must be taken into account for a complete understanding of the
behaviour of all Ce-based HF compounds. We emphasise the heavy
fermion nature of these superconductors; those Ce compounds in
which the f-electrons play little role in their superconductivity
do not concern us for the purposes of this discussion.

We will start by a brief description of the superconducting
pressure-temperature phase diagram of all known Ce-based HF
superconductors, followed by a discussion of the first order
valence transition found in pure Cerium and a few other of its
compounds. After reviewing the theoretical basis for CVF-mediated
superconductivity, we will look at one compound in particular,
\ce, and show that there is very strong evidence for such a
mechanism in this system, and in its isoelectronic sister compound
\cege. Finally, we will discuss to what extent there is evidence
for similar behaviour in other compounds.

\section{Pea versus Potato}

\begin{figure}[th]
\begin{center}
\includegraphics[width=0.9\linewidth]{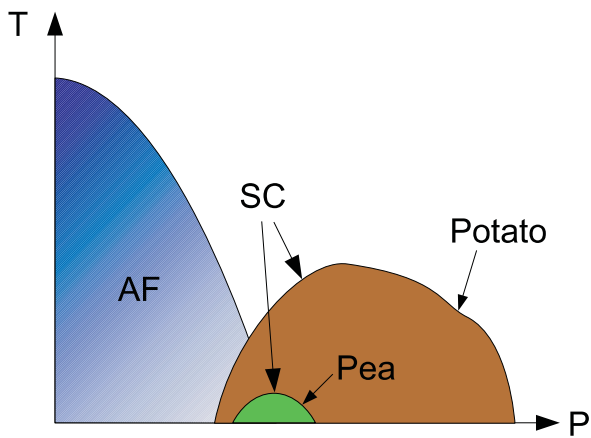}
\end{center}
\caption{Schematic pressure-temperature phase diagrams of two
classes of Ce-based HF superconductors. The `peas' have a small
pocket of superconductivity centred on \pc\ (where
$T_N\rightarrow0$), while the `potatoes' have a large and
asymmetrical superconducting region.} \label{fig:vegetables}
\end{figure}

We can divide Ce-based HF superconductors into two categories,
characterised by the shape of their superconducting region in the
pressure-temperature phase diagram. We will call these two
categories `pea' and `potato' (Fig.~\ref{fig:vegetables}).

The first category contains those which show a small dome of
superconductivity, typically with a pressure width of less than or
equal to 1$\:$GPa, symmetrically situated around the pressure \pc\
where the N\'eel temperature \tn\ tends to zero.  This type of
phase diagram is expected in a spin fluctuation mediated scenario,
where the spin susceptibility diverges at the critical pressure
\pc, and the coherence length $\xi$ of the Cooper pairs increases
with $\left| P-P_{\rm c}\right|$. When $\xi$ is less than the mean
free path $\ell$, \sy\ can occur, leading to a dome-shaped \sg\
region around \pc\ (small and round, like a pea).

In the second category, the `potatoes', are compounds whose \sg\
pressure domains are much less regularly shaped, typically having
a higher maximum \tc, which may not be situated exactly at \pc.
Superconductivity in these compounds can be found over a much
broader pressure range than in the previous category, often
extending far from \pc.

\begin{table*}[tb]
\caption{Classification of Ce-based HF superconductors as
indicated in text. Compounds marked * show traces of
superconductivity. N.b. there is insufficent pressure data to
classify Ce$_2$Ni$_3$Ge$_5$ and Ce$_2$CoIn$_8$.}
\label{Tab:PeaPotato}\vspace{0.2cm}
\begin{center}
\begin{tabular}{cc}
\hline\hline
`Pea' & `Potato'  \\
\hline CeIn$_3$\cite{Mathur98}, CePd$_2$Si$_2$\cite{Mathur98},&
\ce\cite{Holmes04a}, CeCu$_2$Ge$_2$\cite{Jaccard99},CeNi$_2$Ge$_2$\cite{Grosche00},\\
 CeRh$_2$Si$_2$\cite{Movshovich96}, CeCu$_5$Au*\cite{Wilhelm00}&CeCoIn$_5$\cite{Sidorov02},
CeRhIn$_5$\cite{Muramatsu03},
CeIrIn$_5$\cite{Muramatsu03},\\
CeCu$_2$*\cite{Vargoz96} & Ce$_2$RhIn$_8$\cite{Nicklas02}, CeNiGe$_3$\cite{Kotegawa06},\\
 & CePt$_3$Si\cite{Nicklas04}, CeRhSi$_3$\cite{Kimura05}, CeIrSi$_3$\cite{Sugitani06}\\
\hline\hline
\end{tabular}\end{center}

\end{table*}

Table \ref{Tab:PeaPotato} shows the classification of the known
Ce-based HF superconductors for which sufficient data under
pressure exist. Remarkably enough, the `potatoes' outnumber the
`peas' by quite a considerable margin. The phase diagrams were
mostly determined by resistivity.  In CePd$_2$Si$_2$ for example,
bulk superconductivity was found by specific heat measurements
only at exactly \pc \cite{Demuer02}.

In the spin-fluctuation scenario, the scale of \tc\ is set by the
characteristic spin-fluctuation temperature $T_{sf}$. In addition,
it has been suggested with respect to the Ce115 compounds
CeCoIn$_5$, CeRhIn$_5$, and CeIrIn$_5$, that the enhancement of
\tc\ compared to CeIn$_3$ is due to the quasi-two dimensional
character of these compounds\cite{Nicklas01, Monthoux99,
Monthoux01}. Considerations of dimensionality are almost certainly
very important, but they do not resolve the question of the nature
of the pairing mechanism, since the effect applies both to
magnetic and density fluctuation mediated superconductivity
\cite{Monthoux04}.

We emphasise again that in seeking an explanation for the
irregularly shaped superconducting regions found in the majority
of Cerium-based HF superconductors, magnetic fluctuations and
dimensionality are not the only items on the menu. Before
considering a purely magnetic scenario, one must at least rule out
the influence of critical valence fluctuations.

\section{Local versus Itinerant Cerium f Electron}
 The behaviour of Ce and its compounds is dominated by its
single 4f electron. It is rather loosely bound, and can easily be
promoted into the conduction band leaving a non-magnetic Ce$^{4+}$
ion, which has a smaller ionic radius than the Ce$^{3+}$ ion.
Pressure therefore tends to favour a Ce$^{4+}$ configuration.

In the Ce$^{3+}$ (trivalent) state the f electron is localised and
carries a magnetic moment, which can order magnetically, or be
screened by conduction electrons via the Kondo effect. In the
trivalent state, pressure increases the coupling $J$ between the f
and conduction electrons, which tends both to promote the
formation of a Kondo-Yosida singlet, and to encourage magnetic
ordering via the RKKY effect. The Kondo effect increases more
rapidly with $J$, so the overall effect is for magnetic order to
be eventually suppressed, leading to the aforementioned magnetic
QCP where the ordering temperature reaches zero.

Ce compounds at a given pressure can be classified by the nature
of the f-electron into two categories, often referred to as the
Kondo and valence fluctuation regimes respectively:

\hspace{.5cm}
\begin{enumerate}
\begin{minipage}[center]{0.8\columnwidth}
    \item Ce$^{3+}$, 4f$^1$ localised.  Magnetic moment possibly ordered and/or (partially) screened
    by Kondo effect. Kondo temperature $T_K\sim 10\:$K. Properties corresponding to a strongly
    correlated electron system. Kadowaki-Woods ratio\cite{Kadowaki86}
    $A/\gamma^2\simeq 10^{-5}[\mu\Omega$cm mol$^2$K$^2$/J$^2]$.\end{minipage}
\begin{minipage}[center]{0.8\columnwidth}
    \item Ce$^{(3+\delta)+}$,
    4f$^1\rightleftharpoons$4f$^0$+[5d6s], itinerant. $T_K$
    larger than or comparable to crystalline electric field (CEF) splitting. No magnetic order. Intermediate valence.  Kadowaki-Woods ration
    $A/\gamma^2\simeq 0.4\times 10^{-6}[\mu\Omega$cm mol$^2$K$^2$/J$^2]$.\end{minipage}
\end{enumerate}
\vspace{0.5cm}

\subsection{Ce $\alpha-\gamma$ transition}
Cerium metal is well known for the first-order isostructural
volume discontinuity, known as the $\alpha-\gamma$ transition, in
its pressure-temperature phase diagram. This corresponds to a
delocalisation of the Ce 4f electron, as described above. In pure
Ce, the transition itself is strongly first order, with a large
hysteresis, and a critical end point at rather high temperature,
around 600$\:$K and 20$\:$kbar, with a large experimental
uncertainty due to the nature of the
transition\cite{Koskenmaki79}.

%
\subsection{First order transition in cerium compounds}
There are a small number of Ce compounds which also show a first
order valence transition, including CeNi\cite{Gignoux85},
CeP\cite{Jayaraman76}, Ce0.74Th0.26\cite{Shapiro77}
Ce(Rh$_{0.69}$Ir$_{0.31}$)Ge\cite{GaudinE2005}. See also
ref.~\citen{Svane01} and refs therein for a discussion of first
order transitions in Ce, Yb and Eu chalcogenides and pnictides.
These compounds are all notable for containing a high atomic
percentage of Ce, and hence a small Ce-Ce interatomic separation.
None of them display heavy fermion superconductivity.

\section{Theory of Valence Transition and Critical-Valence-Fluctuation
Mediated Superconductivity} \label{sec:theory}
 The idea of CVF-mediated superconductivity and related anomalous phenomena can be backed up
by a microscopic calculations on the basis of a generalized
periodic Anderson model (GPAM)\cite{Onishi00}:
\begin{eqnarray}
H_{\rm GPAM} &=&\sum_{k \sigma}(\epsilon_k-\mu) c_{k
\sigma}^{\dagger}c_{k \sigma}^{} +\varepsilon_{\rm f} \sum_{k
\sigma}f_{k \sigma}^{\dagger}f_{k \sigma}^{} \nonumber
\\
& & +V\sum_{k \sigma}(c_{k \sigma}^{\dagger}f_{k \sigma}^{}+{\rm
h.c.}) +U_{{\rm ff}}\sum_i n_{i \uparrow}^{\rm f} n_{i
\downarrow}^{\rm f} \nonumber
\\
& &+U_{\rm fc}\sum_{i \sigma \sigma'}n_{i \sigma}^{\rm f}n_{i
\sigma'}^{\rm c}, \label{eq:PAMUfc}
\end{eqnarray}
where notations are conventional except for $U_{\rm fc}$, the
local Coulomb repulsion between f- and conduction electorns, which
is a crucial ingredient for a sharp valence transition.

\subsection{Critical-valence-fluctuation mediated superconductivity}
\begin{figure}[th]
\begin{center}
\includegraphics[width=0.9\linewidth]{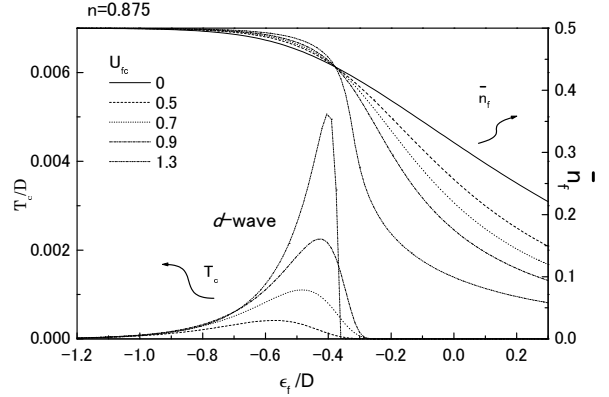}
\caption{ $T_{\rm c}$ for the $d$-wave channel and ${\bar n}_{\rm
f}$, f-electron number per site and ``spin", as a function of
$\epsilon_{\rm f}$. The total number of electrons per site and
``spin" is set as $n=0.875$, and the c-f hybridization is set as
$V=0.5D$. The unit of energy is given by $D$, the Fermi energy of
parabolic conduction band if it were decoupled from f-electrons:
the dispersion $\epsilon_k=k^2/2m-D$ is adopted for the conduction
electrons. } \label{fig:Tcvsef}
\end{center}
\end{figure}

In ref.~\citen{Onishi00}, the model Hamiltonian (\ref{eq:PAMUfc})
with a spherical conduction band (i.e., $\epsilon_{k}=k^{2}/2m-D$,
where $D$ is the Fermi energy of conduction electrons if it were
decoupled from f-electrons) was treated by the mean-field (MF)
approximation in the slave bosons formalism (with $U_{{\rm ff}}\to
\infty$) to discuss the possibility of valence transition, and by
the Gaussian fluctuation theory around the mean-field solution to
discuss a possible superconducting state.  As shown in Fig.~\
\ref{fig:Tcvsef}, the central results of ref.~\citen{Onishi00} are
summarized as follows:
\begin{enumerate}
\def\labelenumi{\theenumi)}
\item Sharp valence change is caused by the effect of $U_{\rm fc}$
with moderate strength, of the order of the Fermi energy ($D$) of
conduction electrons, when the f-level $\epsilon_{\rm f}$ is tuned
to mimic the effect of the pressure. \item The superconducting
state is induced by the process of exchanging slave-boson
fluctuations for the values of $\epsilon_{{\rm f}}$ around which
the sharp valence change occurs. \item The symmetry of the induced
superconducting state is $d$-wave if a spherical model is adopted
for conduction electron.  However, as seen in the argument below,
anisotropic pairing is induced by the CVF modes due to their {\it
almost local} nature.
\end{enumerate}

The peak of $T_{\rm c}$ occurs at $\epsilon_{\rm f}$ slightly
smaller than that corresponding to the steepest slope of ${\bar
n}_{\rm f}$ vs $\epsilon_{\rm f}$ relation, where ${\bar n}_{\rm
f}$ denotes f-electron number per site and ``spin". Since
$\epsilon_{\rm f}$ simulates the pressure variation, this aspect
of the pressure dependence of $T_{\rm c}$ reproduces well that
observed experimentally in CeCu$_2$Ge$_2$ under
pressure~\cite{Jaccard99}. It is also remarked that a sharp change
of ${\bar n}_{\rm f}$ is related to that of the mass enhancement
$m^{*}/m$ by a canonical relation in strongly correlated
limit:\cite{RiceUeda,Shiba}
\begin{equation}
{m^{*}\over m} ={1-{\bar n}_{\rm f}\over 1-2{\bar n}_{\rm f}},
\label{eq:RUS}
\end{equation}
which in turn implies a drastic decrease of the coefficient of the
$T^{2}$-term of the resistivity $A$, because $A$ is proportional
to $\gamma^{2}\propto (m^{*}/m)^{2}$~\cite{Kadowaki86,MMV}.

The $\epsilon_{\rm f}$-dependence of ${\bar n}_{\rm f}$ is smooth
without $U_{\rm fc}$, while its dependence becomes steep as
$U_{\rm fc}/D$ is increased to a moderate strength of ${\cal
O}(1)$. These results are consistent with a physical picture that
the rapid valence change occurs at the condition $\epsilon_{\rm
f}+U_{\rm fc}n_{\rm c}\approx E_{\rm F}$ (the Fermi energy of the
$f^{0}$-state) where the energy of the $f^{0}$- and $f^{1}$-state
are degenerate, leading to enhanced valence fluctuations. For much
larger values of $U_{\rm fc}/D$ or smaller values of $V$ than
those presented in Fig.~\ref{fig:Tcvsef}, there occurs a
first-order like discontinuous transitions although they are not
shown here. However, the valence change occurs more sharply if we
treat the problem in much more proper approximation on the
extended Gutzwiller variational wave function \cite{Onishi002nd}.

It is remarked that $T_{\rm c}$ can exist only for the $d$-wave
($\ell=2$) channel as far as the channels, $\ell=0,1$ and $2$, are
concerned. There exists a sharp peak of $T_{\rm c}$ at around
$\epsilon_{\rm f}$ where ${\bar n}_{\rm f}$ starts to show a rapid
decrease. Its tendency becomes more drastic as $U_{\rm fc}/D$
increases, making the valence change sharper.  In the region where
the f-electron number ${\bar n}_{\rm f}$ is decreased enough,
$T_{\rm c}$ is strongly suppressed. This suggests that the CVF
associated with the sharp valence change of Ce ions is the origin
of the pairing.

\begin{figure}[ht]
\begin{center}
\includegraphics[width=0.9\linewidth]{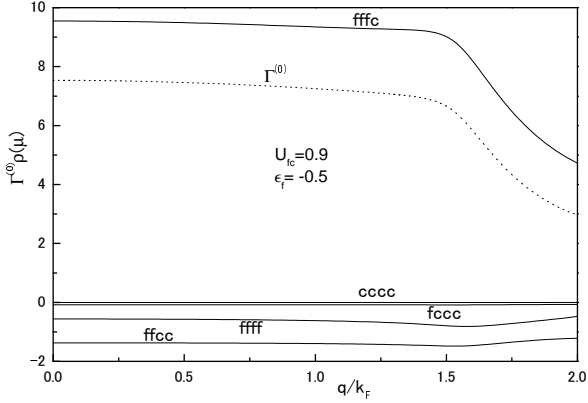}
\caption{Pairing interaction $\Gamma^{0}(q)\rho(0)$, $\rho(0)$
being the quasiparticle density of states at the Fermi
level.\label{fig:PairInt}}
\end{center}
\end{figure}

The pairing interaction $\Gamma^{0}(q)$ ($=V_{{\bf k},{\bf k'}}$
with {\bf q}$={\bf k}-{\bf k}'$) is calculated by taking into
account the Gaussian fluctuations of slave bosons around the MF
solution as mentioned above. The result is shown in Fig.\
\ref{fig:PairInt} for the parameter set $\varepsilon=-0.5D$ and
$U_{\rm fc}=0.9D$ \cite{Onishi00}. One can see clearly that the
scattering process (f,f)$\to$(f,c) or (f,c)$\to$(f,f), in which
the valence of f-electrons is changed, plays a dominant role.  It
is also remarked that $\Gamma^{0}(q)$ is almost $q$-independent up
to $q\simeq (3/2)k_{\rm F}$, reflecting the {\it local} nature of
the valence transition.

\begin{figure}[ht]
\begin{center}
\includegraphics[width=0.9\linewidth]{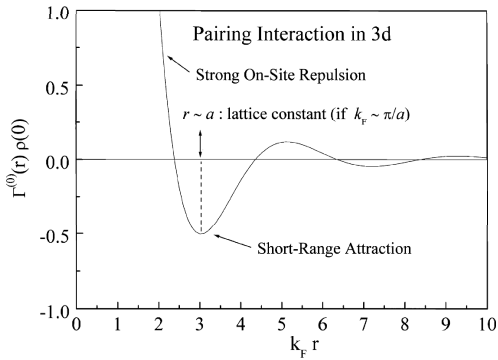}
\caption{Real-space pairing interaction $\Gamma^{0}(r)\rho(0)$.
The parameter set adopted is the same as
Fig.~\ref{fig:PairInt}.\label{fig:PairInt_r}}
\end{center}

\end{figure}

The reason why $d$-wave pairing can be realized is understood in
the following way.  While $\Gamma^{0}(q)$ is always positive and
almost constant at $0<q<(3/2)k_{\rm F}$, giving a short-range
strong repulsion, the sharp decrease at $q>(3/2)k_{\rm F}$ gives
an extended attraction, leading to pairing of non-zero angular
momentum, such as $\ell=1$ and $2$.  One can see this more vividly
by inspecting a real space picture of the pairing interaction
\begin{equation}
\Gamma^{0}(r)=\sum_{{\bf q}}\Gamma^{0}(q)e^{{\rm i}{\bf
q}\cdot{\bf r}}. \label{eq:11}
\end{equation}
The result for $\Gamma^{0}(r)$ is shown in
Fig.~\ref{fig:PairInt_r} for the same parameter set as Fig.~\
\ref{fig:PairInt}. This clearly shows an existence of the extended
attraction together with the on-site strong repulsion.  If we
assume $k_{\rm F}\sim \pi/a$, $a$ being the lattice constant as in
typical metals, the attraction works between nearest neighbor
sites.  Then, according to the discussion in ref.~\citen{MSRV},
pairing of $d$-wave symmetry is promoted. Pairing with $p$-wave
symmetry is also possible in principle in this case because the
attraction mediated by the valence fluctuation has no spin
dependence.

Quite recently, Watanabe {\it et al.} \cite{Watanabe06} reported
results on the density-matrix-renormalisation-group analysis for
the one-dimensional (1d) version of the model Hamiltonian
(\ref{eq:PAMUfc}) in which the conduction band is given as
\begin{equation}
\epsilon_{k}=-t \sum_{i\sigma}
(c_{i\sigma}^{\dagger}c_{i+1\sigma}+c_{i+1\sigma}^{\dagger}c_{i\sigma}),
\label{1d:1}
\end{equation}
where $t$ is the nearest-neighbour hopping integral.  For an
electron number 7/4 per site, $V/t=0.1$ and $U/t=100$, a first
order valence-transition line in $U_{\rm fc}$-$\epsilon_{\rm f}$
plane has been determined, showing that the region of uniform
phase is stabilized and phase separation is suppressed due to
quantum fluctuations.  By analysis of the exponent of the
long-range behaviour of correlation functions of inter-site
pairing, it has been shown that superconducting correlations
become dominant against charge and spin density wave correlations
near the QCP of the valence transition in the region of uniform
phase. This result supports the overall picture of CVF-mediated
unconventional superconductivity discussed in
ref.~\citen{Onishi00}.

\subsection{$T$-linear resistivity and enhanced Sommerfeld
coefficient}
\label{sec:Tlin}Here we briefly discuss how a
critical behavior of the resistivity and the specific heat arises
around the critical valence transition at $P=P_{\rm v}$. The local
nature of the pairing interaction can be seen also in the CVF
spectrum. By exploiting this nature, the valence fluctuation
propagator $\chi_{\rm v}$ (dynamical valence susceptibility) may
be parameterized as
\begin{eqnarray}
\chi_{\rm v}(q,\omega)&\equiv& {\rm i}\int_{0}^{\infty}{\rm d}t
e^{{\rm i}\omega t} \langle[n_{\rm f}(q,t),n_{\rm f}(-q,0)]\rangle
\label{eq:12}
\\
&=&{K\over \omega_{\rm v}-{\rm i}\omega}, \quad{\hbox{for
$q<q_{\rm c}\sim p_{\rm F}$}} \label{eq:13}
\end{eqnarray}
where $n_{\rm f}(q)$ is the Fourier component of the number of
f-electrons per Ce site, and the parameter $\omega_{\rm v}$
parameterizes the closeness to criticality. $\omega_{\rm v}$ is
inversely proportional to the valence susceptibility $\chi_{\rm
v}(0,0)= -(\partial n_{\rm f}/\partial \varepsilon_{\rm
f})_{\mu}$.

\begin{figure}[ht]
\begin{center}
\includegraphics[width=0.4\linewidth]{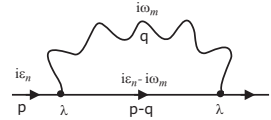}
\caption{Feynman diagram for the self-energy due to
one-fluctuation mode exchange process.  $\lambda$ denotes the
coupling between the valence-fluctuation mode and the
quasiparticles. }
\end{center}
\label{fig:Sigma}
\end{figure}

With the use of the propagator (\ref{eq:13}), it is a
straightforward calculation to obtain a self-energy $\Sigma_{\rm
vf}^{\rm R}(p,\epsilon+{\rm i}\delta)$ of quasiparticle due to one
CVF exchange process shown in Fig.~\ref{fig:Sigma}.
It is reduced to simple forms in typical limiting cases as follows: \\
In the case $\epsilon=0$, $0<T\ll\epsilon_{\rm F}$, $\epsilon_{\rm
F}$ being the effective quasiparticle Fermi energy,
\begin{equation}
{\rm Im}\Sigma_{\rm vf}^{\rm R}(p_{\rm F},0)\simeq
-{|\lambda|^{2}Kq_{\rm c}^{2}\over4\pi^{2} v} \left\{
\begin{array}{@{\,}ll}
\bigl({T\over \omega_{\rm v}}\bigr)^{2}, &T\ll\omega_{\rm v},\\
{\pi\over 2}{T\over \omega_{\rm v}}, &T\gg\omega_{\rm v},
\end{array}
\right. \label{chiv5b}
\end{equation}
where $p_{\rm F}$ is the Fermi momentum, and in the case $T=0$ and
$\epsilon\sim 0$,
\begin{eqnarray}
{\rm Re}\Sigma_{\rm vf}^{\rm R}(p_{\rm F},\epsilon)&\simeq&
-{|\lambda|^{2}Kq_{\rm c}^{2}\over2\pi^{2} v_{\rm
F}}{\epsilon\over\omega_{\rm v}} \int_{0}^{1}{\rm d} u
{1-u^{2}\over u^{2}+1} \ln\biggl|{1\over u }\biggr| \nonumber
\\
&\propto&-{\epsilon_{\rm F}\over \omega_{\rm v}}\epsilon,
\label{chiv7a}
\end{eqnarray}
where $v_{\rm F}$ is the Fermi velocity of quasiparticles, and
$K\sim\epsilon_{\rm F}\rho(0)$ has been used.

The result of eq.~(\ref{chiv5b}) for $T\gg\omega_{\rm v}$ implies
that almost all the critical valence-fluctuation modes can be
regarded as classical at $T>\omega_{\rm v}$, leading to the
$T$-linear resistivity, because the quasiparticles are subject to
large angle scattering by the CVF modes which are effective in a
wide region in the Brillouin zone through the Umklapp process.

Moreover, eq.~(\ref{chiv7a}) implies that an extra enhancement of
the Sommerfeld coefficient is expected other than that of
quasiparticles.  Namely, $\gamma/{\bar \gamma}\propto
\epsilon_{\rm F}/\omega_{\rm v}$ where ${\bar \gamma}$ is the
Sommerfeld coefficient enhanced by the local correlation given by
eq.\ (\ref{eq:RUS}). ${\bar \gamma}$ is a sharply decreasing
function of the pressure (or $\varepsilon_{\rm f}$) around
$P=P_{\rm v}$ while a factor $1/\omega_{\rm v}$ has a sharp peak
at $P=P_{\rm v}$.  Then the resultant $\gamma$ at low temperature
should exhibit a sharp structure around $P=P_{\rm v}$.

\subsection{Enhanced residual resistivity}\label{sec:rho0}
\begin{figure}[ht]
\begin{center}
\includegraphics[width=0.95\linewidth]{\imgpath{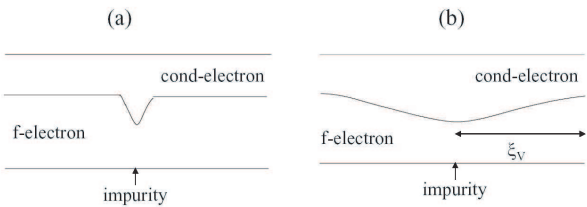}}
\caption{Schematic view of charge distribution of f- and
conduction electrons around impurity: (a) at far from $P\sim
P_{\rm v}$ where the effect of impurity remains as short-ranged so
that the residual resistivity $\rho_{0}$ is not enhanced; (b) at
around $P\sim P_{\rm v}$ where the effect of impurity extends to
long-range region, because the correlation length $\xi_{\rm v}$ of
valence fluctuations diverges as $P\rightarrow P_{\rm v}$, leading
to highly enhanced $\rho_{0}$. } \label{fig:residual}
\end{center}
\end{figure}

The CVF can give rise to huge enhancement of the residual
resistivity $\rho_{0}$ at around $P\sim P_{\rm v}$ through a
many-body effect on the impurity potential.  In the forward
scattering limit, this enhancement is proportional to the valence
susceptibility $-(\partial n_{\rm f}/\partial \epsilon_{\rm
f})_{\mu}$, where $\epsilon_{\rm f}$ is the atomic f-level of the
Ce ion, and $\mu$ is the chemical potential~\cite{Miyake02}.
Physically speaking, local valence change coupled to the impurity
or disorder gives rise to the change of valence in a wide region
around the impurity which then scatters the quasiparticles quite
strongly, leading to the increase of $\rho_{0}$ (see Fig.\
\ref{fig:residual}). Thus the enhancement of $\rho_{0}$ can be
directly related to the degree of sharpness of the valence change
because the variation of the atomic level $\epsilon_{\rm f}$ is
considered to be a smooth function of the pressure. The critical
pressure $P_{\rm v}$ is indeed defined by the maximum of $\rho_0$.

It is remarked that there are two kinds of impurity potential for
quasiparticles consisting mainly of f-electrons: One is due to the
disorder of non-f elements and another is due to the defect of Ce
ions.  The former gives essentially the Born scattering, while the
latter causes the scattering in the unitarity limit.  Then the
residual resistivity is expressed as
\begin{equation}
\rho_{0}=\rho_{0}^{\rm Born}+\rho_{0}^{\rm unit}, \label{residual}
\end{equation}
where $\rho_{0}^{\rm Born}$ is subject to huge enhancement by the
CVF and $\rho_{0}^{\rm unit}$ is essentially unaffected. Then,
eq.~(\ref{residual}) is expressed as
\begin{equation}
\rho_{0}=Bc_{\rm imp}|u(0)|^{2}\ln\biggl|\biggl(-{\partial n_{\rm
f}\over
\partial\epsilon_{\rm f}}\biggr)_{\mu}\bigg/N_{\rm F}\biggr|
+\rho_{0}^{\rm unit} \label{residual2}
\end{equation}
where $c_{\rm imp}$ is the concentration of impurities with
scattering potential $u(q)$, and the coefficient $B$ depends on
the band structure of the host metals. The first term of
eq.~(\ref{residual2}) exhibits a huge enhancement at the critical
valence transition point where $|(\partial n_{\rm
f}/\partial\epsilon_{\rm f})_{\mu}|$ diverges. This huge
enhancement should be compared to the moderate enhancement around
the magnetic quantum critical point where the enhancement arises
only through the renormalization amplitude $z$, as discussed in
ref.~\citen{Miyake02b}.

\section{CeCu$_2$Si$_2$}
The first heavy fermion superconductor to be discovered,
CeCu$_2$Si$_2$, is the archetypical `potato'.  It has a very
irregularly shaped superconducting region under pressure, with a
large enhancement of \tc\ at around 3$\:$GPa, well away from the
antiferromagnetic QCP, which is thought to be at a small positive
pressure $p_c$ of  approximately 0.1$\:$GPa \cite{Gegenwart98}.
\ce\ has been the subject of intensive study by the authors, in
which we showed that the enhancement of \tc\ under pressure was
linked to a sharp Ce valence
transition\cite{Holmes03a,Holmes04a,Jaccard05,Holmes05a,Holmes05b,Holmes06,HolmesPhD}.

The link between superconductivity and valence change had been
noted by Jaccard and coworkers since the first measurements of
\ce\ under high pressure in 1984 \cite{Bellarbi84}. The
isostructural sister compound CeCu$_2$Ge$_2$ was shown to behave
in a very similar way, with a shift of around 10$\:$GPa
corresponding to the larger atomic size of
Ge\cite{Jaccard97,Jaccard99}.

Strong evidence for a sharp valence transition in CeCu$_2$Ge$_2$
was presented by Vargoz \emph{et al.}\cite{VargozPhD,Jaccard99}.
They showed that the $A$ coefficient of a Fermi-liquid fit to the
resistivity $\rho=\rho_0+AT^2$, when plotted against the
low-temperature maximum in resistivity $T^{\rm max}_1$ showed the
expected relationship $A\propto(T^{\rm max}_1)^{-2}$ in two
regions, with a rapid crossover between them.  As $T^{\rm
max}_1\propto T_K \propto \gamma^{-1}$, where $\gamma$ is the
electronic specific heat coefficient, this corresponds to a
crossover from the strongly correlated to weakly correlated branch
of the Kadowaki-Woods plot\cite{MMV}. The maximum in \tc\ seemed
to correspond to the position of this crossover.

Motivated by these results, and by theoretical predictions by
Miyake and coworkers\cite{Miyake99,Onishi00}, we (ATH and DJ)
carried out detailed measurements of resistivity and specific heat
(by the ac calorimetry method) on a single crystal sample of \ce\
in a solid helium pressure medium up to nearly
7$\:$GPa\cite{Holmes04a}.  The use of helium ensures the most
hydrostatic possible conditions at low temperature, a feature of
crucial importance given the rapidly changing behaviour of \ce\
with pressure. Simultaneous measurements of resistivity and
specific heat enabled us to verify the bulk nature of the
superconductivity in the high pressure region, and identify the
most reliable criterion for defining \tc.

We identified a pressure \pv, at about 4.5$\:$GPa, which
corresponds to a valence critical pressure.  Around this point
there are a series of anomalies associated with a delocalisation
of the Ce 4f electron.  \pv\ is slightly above the pressure where
\tc\ reaches a maximum, according to both the theoretical
prediction and our observations.

\subsection{Indications of a Valence Instability}

\begin{figure}[ht]
\begin{center}
\includegraphics[width=0.9\linewidth]{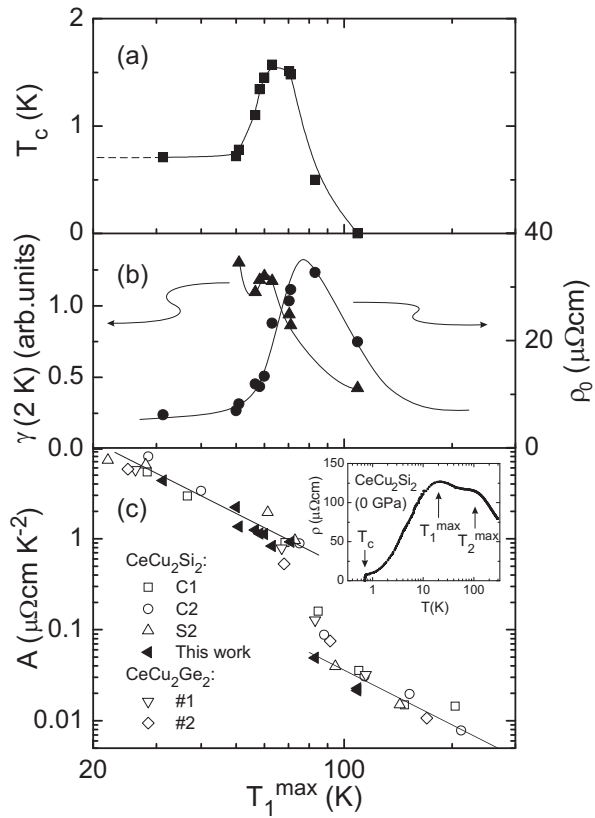}
\end{center}
\caption{Plotted against $T_1^{\rm{max}}$ (defined in
    inset),
     a measure of the characteristic energy scale of the system,
     are (a) the bulk superconducting transition temperature, (b) the
     residual resistivity and estimate $\tilde\gamma$ of the
     Sommerfeld coefficient, and (c) the coefficient $A$ of the
     $\rho\sim A T^2$ law of resistivity, including data from \cege.  Note the straight lines
     where the expected $A\propto (T_1^{\rm{max}})^{-2}$ scaling is
     followed.  The maximum of $T_c$ coincides with the start of
     the region where the scaling relation is broken, while the
     maximum in residual resistivity is situated in the middle of
     the collapse in $A$.  Pressure increases towards the right-hand side
     of the scale (high $T_1^{\rm{max}}$). After ref.~\citen{Holmes04a}.}
\label{fig:Tmax}
\end{figure}

The drastic decrease of the $A$ coefficient of resistivity, and
its change of scaling with respect to $(T^{\rm max}_1)^{-2}$ are
clear signs of the transition from a strongly correlated to weakly
correlated regime.  Table \ref{tab:properties} shows the complete
list of anomalies found around the valence instability in \ce\ and
\cege. Fig.~\ref{fig:Tmax} shows how the enhancement of \tc, and
the anomalies in \rz\ and $\gamma$ correspond to the change from a
strongly to weakly correlated regime.

\begin{table*}[tb]
\caption{\label{tab:properties} Anomalies in \ce\ and \cege\
associated with valence
transition, with references. Symbols explained in the text.\\
Part \textbf{(i)}: Direct evidence for sudden valence change.\\
Part \textbf{(ii)}: Anomalies explained theoretically by the GPAM\cite{Onishi00,Miyake02,Holmes04a} ($\S$\ref{sec:theory}).\\
Part \textbf{(iii)}: Other anomalies observed around crossover to
intermediate valence with pressure.}
\begin{center}
\setlength{\extrarowheight}{2pt}
\begin{tabular}{p{20pt}lcc}
\hline\hline
    &           & \ce&\cege\\
    &           & Ref. & Ref.\\
    \hline
    \multirow{4}{*}{\textbf{(i)}}&Volume discontinuity&-&\citeonline{Onodera02}\\
    &$L_{III}$ X-ray absorption&\citeonline{Roehler88}&-\\
    &Drastic change of $A$ by two orders of magnitude&\citeonline{Holmes04a},\citeonline{Jaccard99}&\citeonline{Jaccard99} \\
     &Change of $A \propto(T_1^{\rm{max}})^{-2}$ scaling &\citeonline{Holmes04a},\citeonline{Jaccard99}&\citeonline{Jaccard99} \\
    \hline
    \multirow{4}{*}{\textbf{(ii)}} &Maximum in $T_c(P)$&\citeonline{Holmes04a},\citeonline{Bellarbi84}&\citeonline{Vargoz98b}\\
    &Large peak in $\rho_0$          &\citeonline{Holmes04a},\citeonline{Jaccard99}
&\citeonline{Jaccard99} \\%
    &Maximum in $\gamma \simeq (C_P/T)$&\citeonline{Holmes04a},\citeonline{Vargoz98}&-\\%
    &$\rho\propto T^n$ from $T_c<T<T^*$, with $n(P_{\rm v})=1$ minimum&    \citeonline{Holmes04a,Bellarbi84,Jaccard85}&\citeonline{Jaccard99}  \\
   \hline
     \multirow{4}{*}{\textbf{(iii)}}    &Sample dependence of $T_c$ & \citeonline{Holmes04a,Holmes05b,Bellarbi84,Thomas96,Thomasson98,Jaccard98,Jaccard99,VargozPhD}
&\citeonline{Jaccard99} \\
     &Enhanced  $\left.\frac{\Delta C_P}{\gamma T}\right|_{T_c}$&\citeonline{Holmes04a}&-\\
    &Resistivity and thermopower indicate $T_1^{\rm{max}}\simeq T_2^{\rm{max}}
$&\citeonline{Jaccard99,Jaccard85}&\citeonline{Jaccard99,Link96}  \\
    &Broad superconducting transition widths $\Delta T_c$ &\citeonline{Holmes04a,Holmes05b,Bellarbi84}&\citeonline{Jaccard99}\\
\hline\hline
\end{tabular}\end{center}

\end{table*}

\subsection{CeCu$_2$(Si$_{1-x}$Ge$_x$)$_2$}

The existence of a second quantum critical point at high pressure
in \ce\ was given further backing by Yuan \emph{et
al.}\cite{Yuan03}. They doped the pure \ce\ system with Ge
(effectively applying negative pressure and introducing disorder)
and then applied pressure to come back along the pressure/volume
axis. For increasing proportions of Ge, the superconductivity was
weakened, and for large enough $x$ the superconducting region
could be split into two domes, near \pc\ and \pv\ respectively.

\subsection{Valence fluctuation mediated
superconductivity in \ce}
\begin{figure}[ht]
\begin{center}
\includegraphics[width=0.9\linewidth]{\imgpath{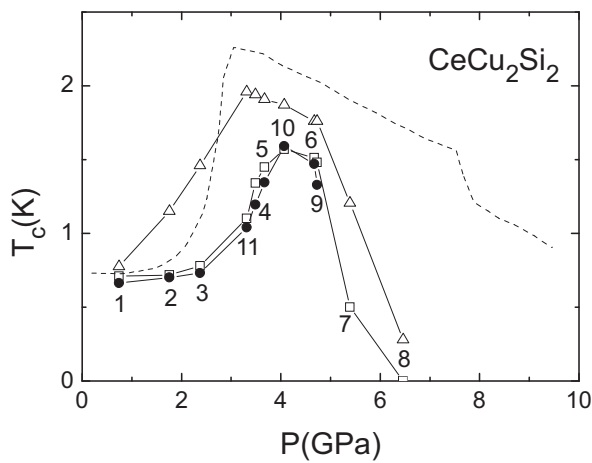}}
\end{center}
\caption{$T_c(P)$ in CeCu$_2$Si$_2$ determined from resistivity
and specific heat measurements. The triangles show $T_c$
determined from the onset of the resistive transition
($T_c^{\rm{onset}}$), the squares show its completion
($T_c^{R=0}$), and the filled circles show the midpoint of the
specific heat jump. The numbers indicate the sequence of
pressures. The dotted line shows $T_c$ determined by
susceptibility in a different sample, also in a helium pressure
medium.\cite{Thomas96} After ref.~\citen{Holmes04a}.}
\label{fig:HeTc}
\end{figure}

\begin{figure}[ht]
\begin{center}
\includegraphics[width=0.9\linewidth]{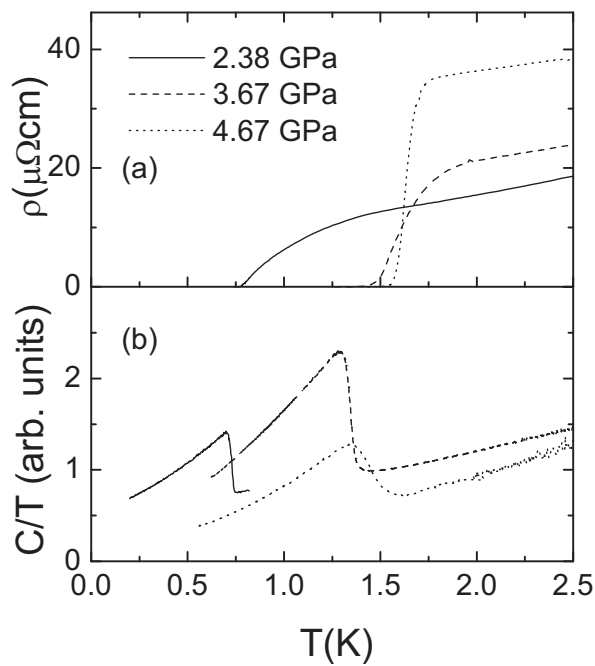}
\end{center}
\caption{Superconducting transition at three pressures in (a)
resistivity and (b) specific heat.  Note the width of the
resistive transitions, and the fact that the start of the jump in
specific heat coincides with the completion of the resistive
transition. After ref.~\citen{Holmes04a}} \label{fig:3Ps}
\end{figure}

Figure \ref{fig:HeTc} shows the pressure variation of the
superconducting transition temperature in \ce\ determined by
specific heat and resistivity. It is clear from this, and
Fig.~\ref{fig:3Ps}, that the bulk transition temperature
determined from the specific heat coincides with the point at
which the resistance reaches zero.  Over a large pressure range
the resistive transitions are very broad. This is often attributed
to pressure gradients in the cell; in this case, however, as a
helium medium was used the gradients should be negligible.

\begin{figure}[ht]
\begin{center}
\includegraphics[width=0.9\linewidth]{\imgpath{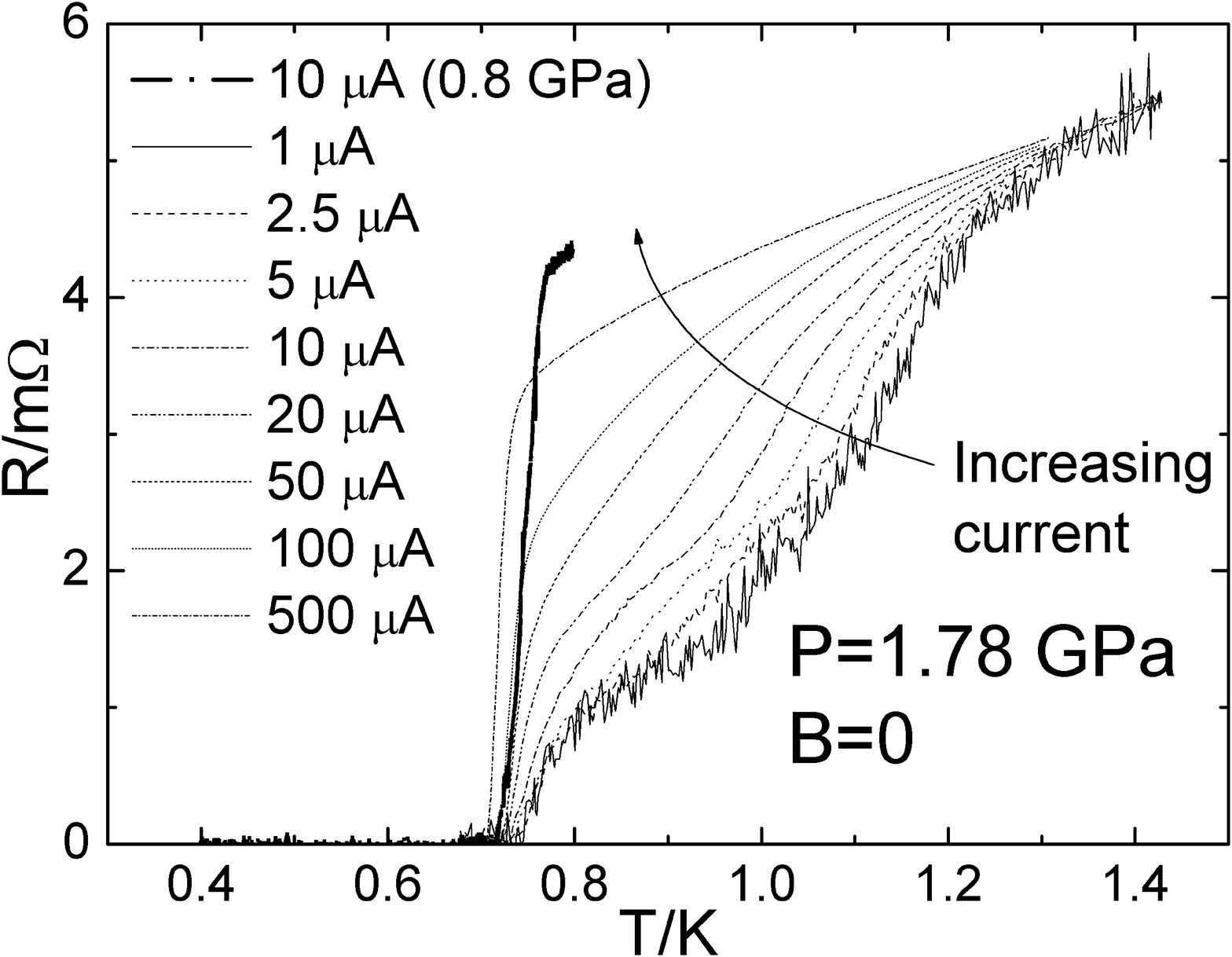}}
\end{center}
\caption{At 1.78$\:$GPa, the resistive transition in \ce\ is very
broad, and appears to have two distinct resistance drops. By
increasing the measurement current, the upper part of the
transition is supressed, and a sharp transition comparable to that
close to ambient pressure is recovered. Note that the pressure
gradient is negligible, so the broad transitions are intrinsic to
the sample.} \label{fig:current}
\end{figure}

\begin{figure}[ht]
\begin{center}
\includegraphics[width=0.9\linewidth]{\imgpath{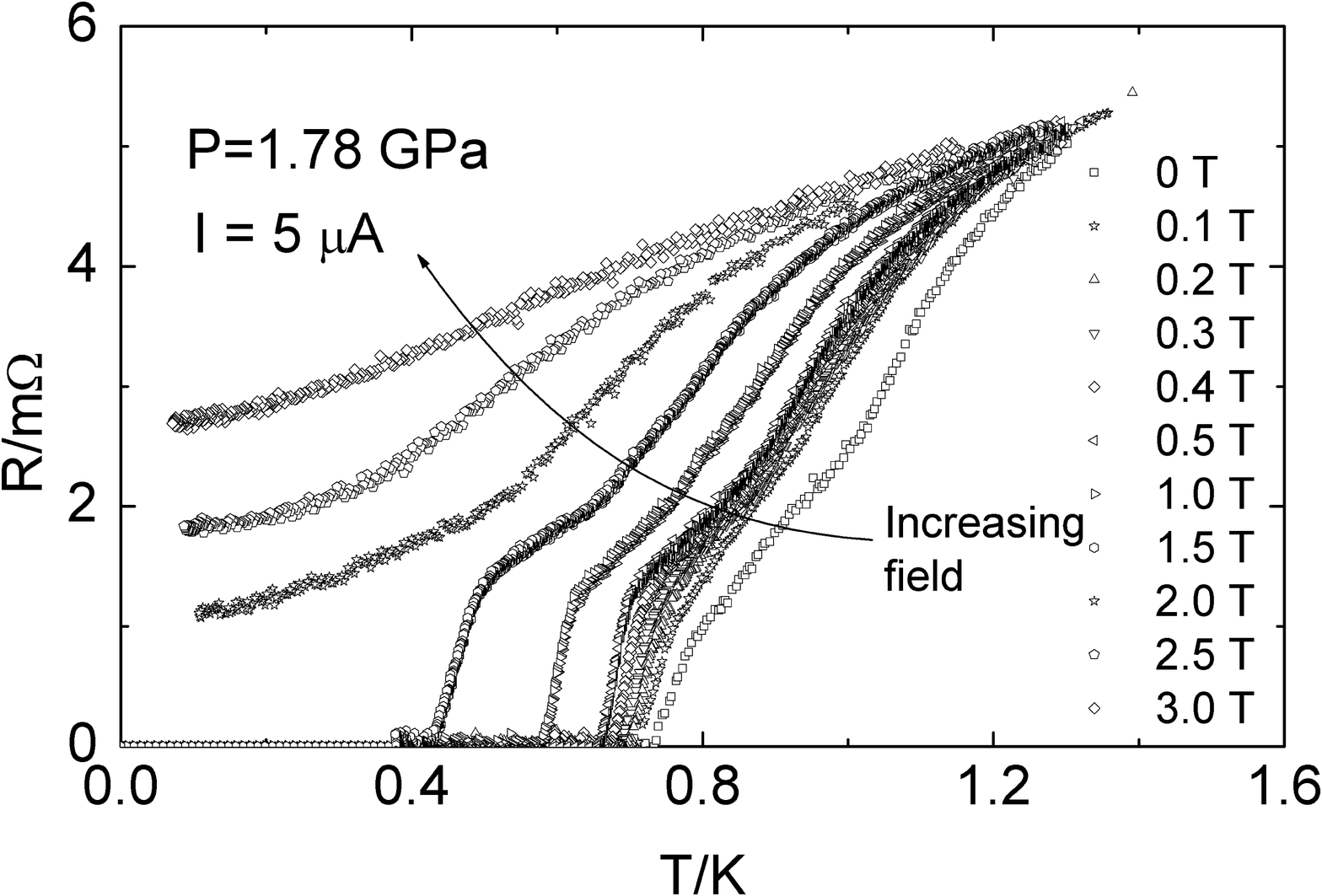}}
\end{center}
\caption{By applying a magnetic field, the lower part of the
resistive transition is supressed more rapidly than the upper
part.} \label{fig:field}
\end{figure}

The nature of the broad resistive transitions was investigated in
detail at 1.78$\:$GPa as a function of measurement current and
magnetic field, as shown in Figs.~\ref{fig:current} and
\ref{fig:field}. Remarkably, the high temperature part of the
resistance drop disappears completely as the current is increased,
recovering a sharp transition.  Applying a magnetic field has the
opposite effect; the low temperature part of the transition is
rapidly suppressed with field, while the upper part remains up to
much higher fields. There appear to be two distinct transitions,
with different \tc's and hence critical fields. The current
dependence implies that the `high \tc' superconductivity is of a
filamentary nature, so its critical current density is easily
exceeded. We should note that even the best samples, with the
highest values of \tc, showed very broad resistive transitions at
certain pressures.

The enhancement of the residual resistivity at \pv\ will be
discussed in more detail below, but it reveals another remarkable
feature of the high pressure superconductivity in \ce. In certain
samples of \ce, the residual resistivity, while relatively large
at ambient pressure, reaches enormous values approaching the
Ioffe-Regel limit around \pv, i.e.~the mean free path is of the
order of the lattice spacing.

\begin{figure}[ht]
\begin{center}
\includegraphics[width=0.9\linewidth]{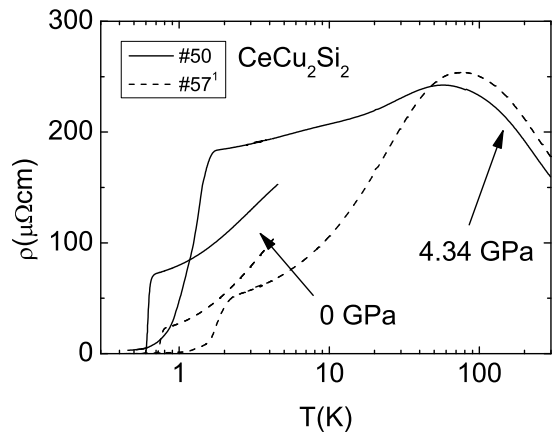}
\end{center}
\caption{Resistivity at ambient pressure and close to \pv\ of two
polycrystalline \ce\ samples prepared by Ishikawa. Note the large
increase in residual resistivity under pressure combined with a
nearly complete resistive transition. After ref.~\citen{Holmes06}}
\label{fig:no5057}
\end{figure}

Magnetically mediated superconductors such as CePd$_2$Si$_2$ are
notoriously dependent on sample quality, superconductivity only
appearing in the purest samples with very small values of \rz, of
the order of a few $\:\mu\Omega$cm.  Fig.~\ref{fig:no5057} shows
two \ce\ samples with very different residual resistivities, in
which a nearly complete resistive transition can be seen at
4.34$\:$GPa despite a residual resistivity of nearly
200$\:\mu\Omega$cm around \pv. This is probably the most
compelling evidence for a new mechanism of superconductivity at
high pressure in \ce.

The superconducting transition temperature and pressure range is
highly sensitive to small uniaxial stresses.  We investigated this
effect by placing single crystal samples from neighbouring
positions on the same source crystal in a Bridgman anvil cell with
their tetragonal $c$-axis parallel and perpendicular to the force
loading direction.  As this technique uses a solid steatite
pressure medium, there remains some non-hydrostatic component to
the stress inside the cell, which can be exploited as a
qualitative control parameter. We showed\cite{Holmes05a} that
there is a significant difference in \tc\ between the two
orientations, and that \pv\ is shifted to a higher pressure in the
case where the force loading direction is perpendicular to the
$c$-axis. This shows from an experimental point of view that any
considerations related to the anisotropy of a system must also be
taken into account in the case of CVF-mediated superconductivity.

\subsection{Normal state properties of \ce\ around \pv}
Three key properties of the normal state in \ce\ show
characteristic behaviour at the valence instability \pv: the
residual resistivity \rz, the power-law behaviour of the
T-dependence of the resistivity, and the Sommerfeld coefficient of
the electronic specific heat capacity $\gamma$. These will be
discussed in the next section.

A key prediction of the GPAM is the enhancement of impurity
scattering around the valence instability discussed in
$\S$\ref{sec:rho0}. This can be thought of as an impurity
nucleating a change of valence in the Ce atoms surrounding it,
thereby inflating its scattering cross-section, and hence the
residual resistivity, as given by eq.~(\ref{residual2}).

\begin{figure}[ht]
\begin{center}
\includegraphics[width=0.9\linewidth]{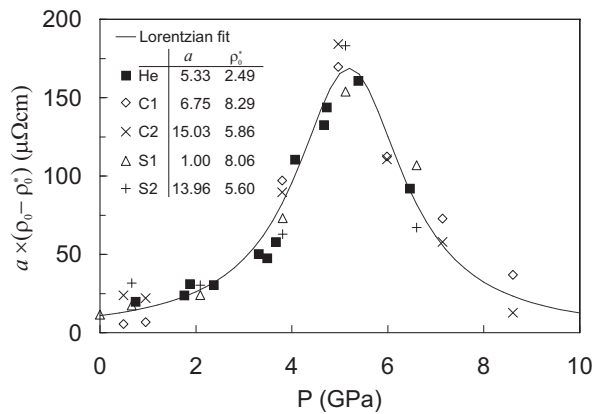}
\end{center}
\caption{Enhancement of residual resistivity in several different
\ce~samples, scaled to a universal pressure dependence, with $a$
and $\rho_0^*$ being normalizing factors. The maximum in $\rho_0$
is at a pressure slightly higher than that corresponding to the
maximum in $T_c$. After ref.~\citen{Holmes04a}.} \label{fig:rho0}
\end{figure}

Fig.~\ref{fig:rho0} shows the pressure dependence of \rz\ in a
number of different \ce\ samples, which can all be scaled on to
the same Lorenzian curve by a simple linear transformation. This
scaling behavior of $\rho_{0}$ would be possible if the universal
form is given by $\ln|(-\partial n_{\rm f}/\partial\epsilon_{\rm
f})_{\mu}/N_{\rm F}|$.

It should be noted that the impurity contribution to the
resistivity has a negative temperature dependence, a fact
especially evident at very high pressure where the $A$-coefficient
of the resistivity becomes small.  Even in samples with a small
\rz\ this can cause problems when fitting power laws to the
resistivity in the region above \pv.

\begin{figure}[ht]
\begin{center}
\includegraphics[width=0.9\linewidth]{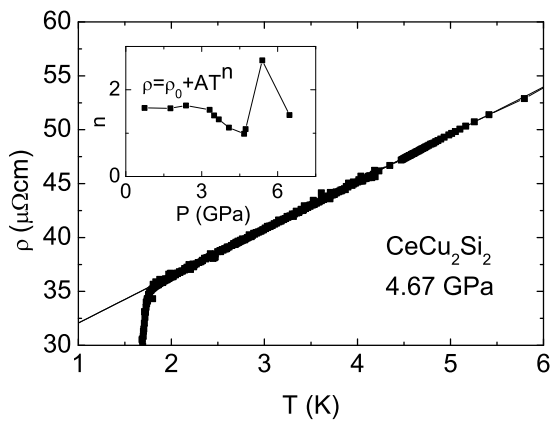}
\end{center}
\caption{Linear resistivity $\rho-\rho_0\propto T$ is found over a
broad temperature range at pressures close to \pv.  The inset
shows the result of a power law fit at different pressures between
\tc\ and 4.2$\:$K.} \label{fig:LinearResist}
\end{figure}

Due to the localised nature of the critical valence fluctuations,
the quasiparticle scattering is nearly $q$-independent, leading to
a linear resistivity, implied by eq.~(\ref{chiv5b}).
Fig.~\ref{fig:LinearResist} shows the resistivity at a pressure
very close to \pv, where a strictly linear temperature dependence
is observed over a broad temperature range.

\begin{figure}[ht]
\begin{center}
\includegraphics[width=0.9\linewidth]{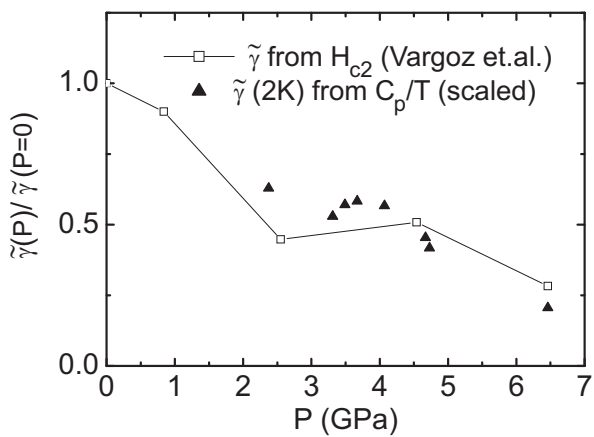}
\end{center}
\caption{Estimate $\tilde\gamma(P)$ of the Sommerfeld coefficient
from a.c.-calorimetry signal at 2$\:$K (triangles), scaled for
comparison with that deduced from $H_{c2}$ measurements
(squares).\cite{Vargoz98} The noise on the calorimetry signal is
smaller than the symbol size, however see ref.~\citen{Holmes04a}
for a discussion of possible systematic errors.} \label{fig:gamma}
\end{figure}

The electronic specific heat coefficient $\gamma$, and hence the
effective mass $m^*/m$, can be estimated by following the
calorimetric signal $C/T$, at a fixed temperature and measurement
frequency above the superconducting transition, though this
includes constant or slowly varying addenda from the helium,
diamonds etc. Figure~\ref{fig:gamma} shows the estimate
$\tilde\gamma(P)$, along with the value deduced from measurements
of the upper critical field\cite{Vargoz98}. A single constant
scale factor has been introduced, showing that the two curves can
be superimposed. There is a clear anomaly in $\tilde\gamma$ at
4$\:$GPa (just below the pressure corresponding to
$T_c^{\rm{max}}$), superimposed on a constant reduction with
pressure. The effective mass is also reflected in the initial
slope of the upper critical field $H_{c2}'(T_c)$, which in our
sample also had a maximum at the same pressure as the peak in
$\tilde\gamma$. Equation (\ref{chiv7a}) implies an enhancement of
$\gamma$ at \pv\ due to critical valence fluctuations; combined
with a monotonic reduction in effective mass due to the rapid
change of $n_{\rm f}$ with pressure, this leads to a peak in
$\gamma$ observed slightly below \pv.

\begin{figure}[ht]
\begin{center}
\includegraphics[width=0.9\linewidth]{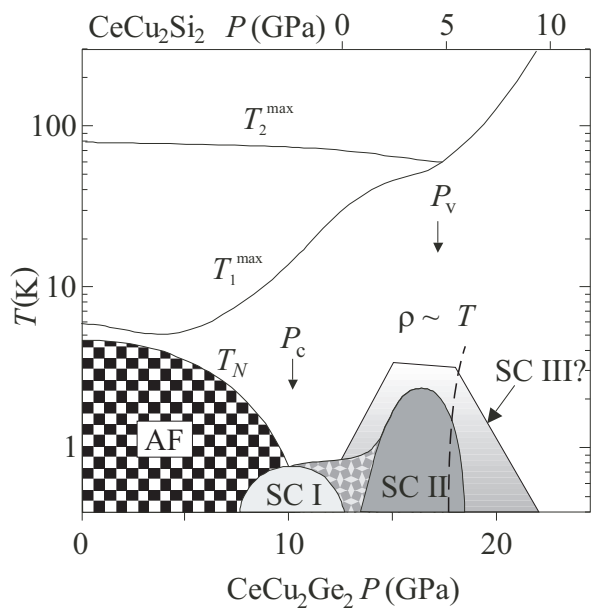}
\end{center}
\caption{Schematic $P$-$T$ phase diagram for CeCu$_2$(Si/Ge)$_2$
showing the two critical pressures
 $P_{\rm c}$ and $P_{\rm v}$. At $P_{\rm c}$, where the antiferromagnetic ordering
temperature $T_N\rightarrow 0$,
  superconductivity in region $SC\ I$ is mediated by
 antiferromagnetic spin fluctuations; around $P_{\rm v}$, in the region $SC\
II$, valence
 fluctuations provide the pairing
 mechanism and the resistivity is linear in temperature.  The
 temperatures $T_1^{\rm{max}}$, and $T_1^{\rm{max}}$,
 merge at a pressure coinciding with $P_{\rm v}$. The dashed line represents a hypothetical (weakly)
 first order valence transition with a critical end point close to zero temperature.} \label{fig:phasediag}
\end{figure}

\section{Other Candidates for Critical Valence Fluctuation Mediated SC}
After \ce\ and \cege, the next strongest candidates for CVF
mediated \sy\ are probably the Ce115 compounds.  Between them they
show a number of the properties discussed above, namely linear
resistivity in CeCoIn$_5$\cite{Petrovic01}, and linear resisity
and a pressure-induced maximum of \rz\ in CeRhIn$_5$
\cite{Hegger00,Muramatsu01}.

CeIrIn$_5$ is probably the strongest candidate for a CVF mediated
superconductor at ambient pressure. Pressure studies on
CeRh$_{1-x}$Ir$_x$In$_5$ have showed that the superconducting
region can be split into two distinct pockets, with a cusp-like
minimum of \tc\ at around $x=0.9$.\cite{nicklas:020505}.
$^{115}$In-NQR measurements of stoichiometric CeIrIn$_5$ under
pressure showed that while \tc\ increased under pressure, reaching
$T_c^{\rm max}\simeq1\:$K at 3$\:$GPa, more than twice its value
at ambient pressure, the nuclear-spin-lattice-relaxation rate
$(1/T_1)$ was found to decrease monotonically with pressure. This
indicates that the antiferromagnetic spin fluctuations are
\emph{decreasing} with pressure, while the superconductivity is
strengthening.

The cases of PuCoGa$_5$\cite{Sarrao02} and
PuRhGa$_5$\cite{Wastin03} are also very intriguing. They become
superconducting at the comparatively high temperatures of 18$\:$K
and 9$\:$K respectively. Pu metal itself has very rich phase
diagram whose origin may be traced back to the fact that the 5f
electrons in Pu are located near the boundary between the
localized state with nearly an integral valence and the itinerant
state with a fractional valence. There exists some circumstantial
evidence to support the possibility of CVF-mediated \sy\ in these
compounds. The ratio $A/\gamma^2$ of the Pu115 compounds
($A/\gamma \sim 2 \times 10^{-6} [\mu\Omega$cm mol$^2$K$^2$/J$^2]$
for PuCoGa5 \cite{WastinPC}) lies between that of strongly and
weakly correlated metals. This suggests that Pu115 compounds are
not strongly correlated metals but depend on the valence
fluctuations of the Pu ion, although the Sommerfeld coefficient
$\gamma\sim 95$[mJ/K$^2$mol] is moderately enhanced.

There are other examples among the `potato' compounds mentioned
above, such as the two separate domes of superconductivity in the
phase diagram of CeNi$_2$Ge$_2$, and broad resistive transitions
and apparent enhancement of \rz\ under pressure in CeNiGe$_3$, but
these have not been studied in so much detail.

In \ce\ and \cege\ the two critical pressures \pv\ and \pc\
corresponding to the magnetic and valence instabilities are well
separated. It may be the case in other systems, however, that
\pv$\sim$\pc, or even \pv$<$\pc. There is a clear need for some
sort of `smoking gun' to distinguish between valence and
magnetically mediated superconductivity. Both are predicted to be
$d$-wave, but knowledge of the precise gap symmetry, though very
difficult to determine under pressure, may be the key to
distinguishing these two mechanism.

\section{Conclusions}
Ce metal and a small number of its compounds show a first order
valence transition in their pressure-temperature phase diagrams.
We contend that an analogous first-order or nearly first order
transition is present in many Ce compounds, and in certain
circumstances, fluctuations around its critical end point can
mediate superconductivity.

\ce\ has been shown almost unambiguously to have a region of
superconductivity at high pressure mediated by such critical
valence fluctuations, close to a valence instability at a pressure
around 4.5$\:$GPa. There are a number of anomalies in the normal
state around this pressure which can be tied to an abrupt
delocalisation of the Ce 4f electron.

A majority of the other known Ce-based HF superconductors have
characteristics which cannot be explained purely in a spin
fluctuation mediated scenario. The presence of a valence
instability in these compounds may be quite general, and the key
to a more complete understanding of their behaviour.


\section*{Acknowledgment}
This work was supported by a Grant-in-Aid for Creative Scientific
Research (15GS0213), a Grant-in-Aid for Scientific Research
(Nos.~16340103 \& 15204032), and the 21st Century COE Program
(G18) by the Japan Society for the Promotion of Science.


\section*{References}
\begin{thebibliography}{10}
\expandafter\ifx\csname url\endcsname\relax
  \def\url#1{\texttt{#1}}\fi
\expandafter\ifx\csname
urlprefix\endcsname\relax\def\urlprefix{URL }\fi

\bibitem{Onishi00}
Y.~Onishi and K.~Miyake: J. Phys. Soc. Jpn.  \textbf{69} (2000)
3955.

\bibitem{Demuer02}
A.~Demuer, A.~T. Holmes and D.~Jaccard: J. Phys: Condens. Matter
\textbf{14}
  (2002) L529.

\bibitem{Nicklas01}
M.~Nicklas, R.~Borth, E.~Lengyel, P.~G. Pagliuso, J.~L. Sarrao,
V.~A. Sidorov,
  G.~Sparn, F.~Steglich and J.~D. Thompson: J. Phys: Condens. Matter
  \textbf{13} (2001) L905.

\bibitem{Monthoux99}
P.~Monthoux and G.~G. Lonzarich: Phys. Rev. B  \textbf{59} (1999)
14598.

\bibitem{Monthoux01}
P.~Monthoux and G.~G. Lonzarich: Phys. Rev. B  \textbf{63} (2001)
054529.

\bibitem{Monthoux04}
P.~Monthoux and G.~G. Lonzarich: Phys. Rev. B  \textbf{69} (2004)
064517.

\bibitem{Mathur98}
N.~D. Mathur, F.~M. Grosche, S.~R. Julian, I.~R. Walker, D.~M.
Freye, R.~K.~W.
  Haselwimmer and G.~G. Lonzarich: Nature  \textbf{394} (1998) 39.

\bibitem{Holmes04a}
A.~T. Holmes, D.~Jaccard and K.~Miyake: Phys. Rev. B  \textbf{69}
(2004)
  024508.

\bibitem{Jaccard99}
D.~Jaccard, H.~Wilhelm, K.~Alami-Yadri and E.~Vargoz: Physica B
  \textbf{259--261} (1999) 1.

\bibitem{Grosche00}
F.~M. Grosche, P.~Agarwal, S.~R. Julian, N.~J. Wilson, R.~K.~W.
Haselwimmer,
  S.~J.~S. Lister, N.~D. Mathur, F.~V. Carter, S.~S. Saxena and G.~G.
  Lonzarich: J. Phys: Condens. Matter  \textbf{12} (2000) L533.

\bibitem{Movshovich96}
R.~Movshovich, T.~Graf, D.~Mandrus, J.~D. Thompson, J.~L. Smith
and Z.~Fisk:
  Phys. Rev. B  \textbf{53} (1996) 8241.

\bibitem{Wilhelm00}
H.~Wilhelm, S.~Raymond, D.~Jaccard, O.~Stockert, H.~v.~Loehneysen
and A.~Rosch: in
  \emph{Proc. AIRAPT-17,
  Hawaii, 1999} ed. M.~Manghnani, W.~Nellis and M.~Nicol (Universities Press, Hyderabad, 2000) p.~697.

\bibitem{Sidorov02}
V.~A. Sidorov, M.~Nicklas, P.~G. Pagliuso, J.~L. Sarrao, Y.~Bang,
A.~V.
  Balatsky and J.~D. Thompson: Phys. Rev. Lett.  \textbf{89} (2002) 157004.

\bibitem{Muramatsu03}
T.~Muramatsu, T.~C. Kobayashi, K.~Shimizu, K.~Amaya, D.~Aoki,
Y.~Haga and
  Y.~Onuki: Physica C  \textbf{388-389} (2003) 539.

\bibitem{Vargoz96}
E.~Vargoza, P.~Link, D.~Jaccard, T.~L. Bihan and S.~Heathman:
Physica B
  \textbf{229} (1996) 225.

\bibitem{Nicklas02}
M.~Nicklas, V.~A. Sidorov, H.~A. Borges, P.~G. Pagliuso,
C.~Petrovic, Z.~Fisk,
  J.~L. Sarrao and J.~D. Thompson: Physical Review B (Condensed Matter and
  Materials Physics)  \textbf{67} (2003) 020506.

\bibitem{Kotegawa06}
H.~{Kotegawa}, T.~{Miyoshi}, K.~{Takeda}, S.~{Fukushima},
H.~{Hidaka},
  K.~{Tabata}, T.~C. {Kobayashi}, M.~{Nakashima}, A.~{Thamizhavel}, R.~{Settai}
  and Y.~{{\= O}nuki}: Physica B  \textbf{378} (2006) 419.

\bibitem{Nicklas04}
M.~{Nicklas}, G.~{Sparn}, R.~{Lackner}, E.~{Bauer} and
F.~{Steglich}: Physica B
   \textbf{359} (2005) 386.

\bibitem{Kimura05}
N.~Kimura, K.~Ito, K.~Saitoh, Y.~Umeda, H.~Aoki and T.~Terashima:
Phys. Rev.
  Lett.  \textbf{95} (2005) 247004.

\bibitem{Sugitani06}
I.~{Sugitani}, Y.~{Okuda}, H.~{Shishido}, T.~{Yamada},
A.~{Thamizhavel},
  E.~{Yamamoto}, T.~D. {Matsuda}, Y.~{Haga}, T.~{Takeuchi}, R.~{Settai} and
  Y.~{{\= O}nuki}: J. Phys. Soc. Jpn.  \textbf{75} (2006) 3703.

\bibitem{Kadowaki86}
K.~Kadowaki and S.~Woods: Solid State Commun.  \textbf{58} (1986)
507.

\bibitem{Koskenmaki79}
D.~G.~Koskenmaki and K.~A.~Gschneidner, Jr.: in \emph{Handbook on
the Physics and Chemistry of Rare Earths}, edited by
K.~A.~Gschneidner, Jr. and L.~Eyring (North-Holland, Amsterdam,
1979), Vol. I, p. 337.

\bibitem{Gignoux85}
D.~Gignoux and J.~Voiron: Phys. Rev. B  \textbf{32} (1985) 4822.

\bibitem{Jayaraman76}
A.~Jayaraman, W.~Lowe, L.~D. Longinotti and E.~Bucher: Phys. Rev.
Lett.
  \textbf{36} (1976) 366.

\bibitem{Shapiro77}
S.~M. Shapiro, J.~D. Axe, R.~J. Birgeneau, J.~M. Lawrence and
R.~D. Parks:
  Phys. Rev. B  \textbf{16} (1977) 2225.

\bibitem{GaudinE2005}
E.~Gaudin, B.~Chevalier, B.~Heying, U.~Rodewald and R.~Pottgen:
Chem. of Mater.
   \textbf{17} (2005) 2693.

\bibitem{Svane01}
A.~Svane, P.~Strange, W.~Temmerman, Z.~Szotek, H.~Winter and
L.~Petit: Phys.
  Stat. Sol. (b)  \textbf{223} (2001) 105.

\bibitem{RiceUeda}
T.~M. Rice and K.~Ueda: Phys. Rev. B  \textbf{34} (1986) 6420.

\bibitem{Shiba}
H.~Shiba: J. Phys. Soc. Jpn.  \textbf{55} (1986) 2765.

\bibitem{MMV}
K.~Miyake, T.~Matsuura and C.~M. Varma: Solid State Commun.
\textbf{71} (1989)
  1149.

\bibitem{Onishi002nd}
Y.~Onishi and K.~Miyake: Physica B  \textbf{281--282} (2000) 191.

\bibitem{MSRV}
K.~Miyake, S.~Schmitt-Rink and C.~M. Varma: Phys. Rev. B
\textbf{34} (1986)
  6554.

\bibitem{Watanabe06}
S.~{Watanabe}, M.~{Imada} and K.~{Miyake}: Journal of the Physical
Society of
  Japan  \textbf{75} (2006) 3710.

\bibitem{Miyake02}
K.~Miyake and H.~Maebashi: J. Phys. Soc. Jpn.  \textbf{71} (2002)
1007.

\bibitem{Miyake02b}
K.~Miyake and O.~Narikiyo: J. Phys. Soc. Jpn.  \textbf{71} (2002)
867.

\bibitem{Gegenwart98}
P.~Gegenwart, C.~Langhammer, C.~Geibel, R.~Helfrich, M.~Lang,
G.~Sparn,
  F.~Steglich, R.~Horn, L.~Donnevert, A.~Link and W.~Assmus: Phys. Rev. Lett.
  \textbf{81} (1998) 1501.

\bibitem{Holmes03a}
A.~T. Holmes, A.~Demuer and D.~Jaccard: Acta Phys. Pol., B
\textbf{34} (2003)
  567.

\bibitem{Jaccard05}
D.~Jaccard and A.~T. Holmes: Physica B  \textbf{359--361} (2005)
333.

\bibitem{Holmes05a}
D.~{Jaccard} and A.~T. {Holmes}: Physica B  \textbf{359} (2005)
333.

\bibitem{Holmes05b}
A.~T. Holmes, D.~Jaccard, H.~S. Jeevan, C.~Geibel and M.~Ishikawa:
J. Phys:
  Condens. Matter  \textbf{17} (2005) 5423.

\bibitem{Holmes06}
A.~T. {Holmes} and D.~{Jaccard}: Physica B Condensed Matter
\textbf{378}
  (2006) 339.

\bibitem{HolmesPhD}
A.~T. Holmes: Dr. Thesis, University of Geneva, no. 3539, Geneva
2004.
\newline\url{http://www.unige.ch\linebreak[4]/cyberdocuments/theses2004/HolmesAT/%
these.pdf}

\bibitem{Bellarbi84}
B.~Bellarbi, A.~Benoit, D.~Jaccard, J.~M. Mignot and H.~F. Braun:
Phys. Rev. B
  \textbf{30} (1984) 1182.

\bibitem{Jaccard97}
D.~Jaccard, P.~Link, E.~Vargoz and K.~Alami-Yadri: Physica B
\textbf{230--232}
  (1997) 297.

\bibitem{VargozPhD}
E.~Vargoz: Dr. Thesis, University of Geneva, no. 3003, Geneva
1998.

\bibitem{Miyake99}
K.~Miyake, O.~Narikiyo and Y.~Onishi: Physica B  \textbf{259--261}
(1999) 676.

\bibitem{Onodera02}
A.~Onodera, S.~Tsuduki, Y.~Ohishi, T.~Watanuki, K.~Ishida,
Y.~Kitaoka and
  Y.~Onuki: Solid State Commun.  \textbf{123} (2002) 113.

\bibitem{Roehler88}
J.~Roehler, J.~Klug and K.~Keulerz: J. Magn. Magn. Mater.
\textbf{76--77}
  (1988) 340.

\bibitem{Vargoz98b}
E.~Vargoz and D.~Jaccard: J. Magn. Magn. Mater.  \textbf{177--181}
(1998) 294.

\bibitem{Vargoz98}
E.~Vargoz, D.~Jaccard, J.~Y. Genoud, J.~P. Brison and J.~Flouquet:
Solid State
  Commun.  \textbf{106} (1998) 631.

\bibitem{Jaccard85}
D.~Jaccard, J.~M. Mignot, B.~Bellarbi, A.~Benoit, H.~F. Braun and
J.~Sierro: J.
  Magn. Magn. Mater.  \textbf{47--48} (1985) 23.

\bibitem{Thomas96}
F.~Thomas, C.~Ayache, I.~A. Fomine, J.~Thomasson and C.~Geibel: J.
Phys:
  Condens. Matter  \textbf{8} (1996) L51.

\bibitem{Thomasson98}
J.~Thomasson, Y.~Okayama, I.~Sheikin, J.~P. Brison and
D.~Braithwaite: Solid
  State Commun.  \textbf{106} (1998) 637.

\bibitem{Jaccard98}
D.~Jaccard, E.~Vargoz, K.~Alami-Yadri and H.~Wilhelm: Rev. High
Pressure Sci.
  Techno.  \textbf{7} (1998) 412.

\bibitem{Link96}
P.~Link, D.~Jaccard and P.~Lejay: Physica B  \textbf{225} (1996)
207.

\bibitem{Yuan03}
H.~Q. Yuan, M.~Deppe, G.~Sparn, C.~Geibel and F.~Steglich: Acta
Phys. Pol., B
  \textbf{34} (2003) 533.

\bibitem{Petrovic01}
C.~Petrovic, P.~G. Pagliuso, M.~F. Hundley, R.~Movshovich, J.~L.
Sarrao, J.~D.
  Thompson, Z.~Fisk and P.~Monthoux: J. Phys: Condens. Matter  \textbf{13}
  (2001) L337.

\bibitem{Hegger00}
H.~Hegger, C.~Petrovic, E.~G. Moshopoulou, M.~F. Hundley, J.~L.
Sarrao, Z.~Fisk
  and J.~D. Thompson: Phys. Rev. Lett.  \textbf{84} (2000) 4986.

\bibitem{Muramatsu01}
T.~Muramatsu, N.~Tateiwa, T.~C. Kobayashi, K.~Shimizu, K.~Amaya,
D.~Aoki,
  H.~Shishido, Y.~Haga and Y.~Onuki: J. Phys. Soc. Jpn.  \textbf{70} (2001)
  3362.

\bibitem{nicklas:020505}
M.~Nicklas, V.~A. Sidorov, H.~A. Borges, P.~G. Pagliuso, J.~L.
Sarrao and J.~D.
  Thompson: Phys. Rev. B  \textbf{70} (2004) 020505.

\bibitem{Sarrao02}
J.~L. Sarrao, L.~A. Morales, J.~D. Thompson, B.~L. Scott, G.~R.
Stewart,
  F.~Wastin, J.~Rebizant, P.~Boulet, E.~Colineau and G.~H. Lander: Nature
  \textbf{420} (2002) 297.

\bibitem{Wastin03}
F.~Wastin, P.~Boulet, J.~Rebizant, E.~Colineau and G.~H. Lander:
J. Phys:
  Condens. Matter  \textbf{15} (2003) S2279.

\bibitem{WastinPC}
F.~Wastin: Private communication.

\end{thebibliography}
\end{document}